\documentclass[aps,preprint]{revtex4}%
\usepackage{amsfonts}
\usepackage{amsmath}
\usepackage{amssymb}
\usepackage{graphicx}%
\begin{document}
\title{Spin-polarized quantum transport through a $T$-shape quantum dot-array: a
model of spin splitter }
\author{Rui Wang, J.-Q. Liang$^{\ast}$}
\affiliation{Institute of Theoretical Physics Shanxi University Taiyuan 030006 China}


\begin{abstract}
We in this paper study theoretically the spin-polarized quantum transport
through a $T$-shape quantum dot-array by means of transfer-matrix method along
with the Green$^{,}$s function technique. Multi-magnetic fields are used to
produce the spin-polarized transmission probabilities and therefore the spin
currents, which are shown to be tunable in a wide range by adjusting the
energy, and the direction-angle of magnetic fields as well. Particularly the
opposite- spin- polarization currents separately flowing out to two electrodes
can be generated and thus the system acts as a spin splitter.

\end{abstract}
\maketitle
\tableofcontents

$\ast$jqliang@sxu.edu.cn


\section{INTRODUCTION}

Recent progress in the nano-fabrication of quantum devices enables us to study
electron transport through quantum dots(QDs) in a controllable way
\cite{Gordon,Cronenwett,Orellana} since a wide range of energy can be achieved
here by continuously changing the applied external potential in contrast with
real atoms. A QD- array regarded as an artificial
crystal\cite{Holleitner,Holleitner2,hangguan} also becomes reliable due to the
recent development of nanotechnology. When the size of structures is small
with respect to the coherence length, the transmission probability plays an
essential role in the quantum transport. For example, we can obtain the
conductance of the devices from the transmission probability with the help of
the Laudauer-B$\ddot{u}$ticker formula. Various methods have been developed
for numerical calculations of the transmission probability
\cite{Kirczenow,Ulloa,Brum,Tamura,Wu,Thouless,Lee,Schweizer,Schweizer2,Mackinnon,Ando,Xu}%
. The quantum transport through a multi-terminal
system\cite{BIGNON,Tran,Texier} can be analyzed in terms of all transmission
probabilities between any pair of terminals in principle. B$\ddot{u}$ticker et
al.\cite{Buttiker} derived a scattering matrix for a three-terminal junction
consisting of homogeneous wires based on the unitary condition and the
assumption that the scattering matrix is real. It was shown that this matrix
is energy independent and the coupling parameter has to be determined from
phenomenological arguments\cite{Mello,Takai}. An energy-dependent scattering
matrix for three-terminal junction was found later \cite{Itoh}.

Spintronics has become a new branch of condensed matter physics and material
science where not only the charge but also the spin degree of freedom of
electrons play an essential role\cite{Z}. It is of fundamental importance to
study the mechanism to generate and control the spin-polarized currents in
spintronic devices which may promise practical applications in quantum
computing and information\cite{S,G}. In recent yeas the spin-polarized
transports in quantum dots have attracted considerable attentions both
theoretically\cite{Y,A,P,Q,B,T,C,K,Qi,H,Dong,Dong2} and
experimentally\cite{Su,M}, where the spin currents are generated either by
magnetic material electrodes or by the rotating-magnetic field which results
in the spin-flip. Most of the studies are concentrated on the one or two-QD
devices with only one output electrode and the spin-polarized transport in
QD-array consisting of arbitrary number of QDs with multi-terminal has not yet
be investigated. In this paper we study the quantum transport through a
$T$-shape QD-array with three electrodes in which the spin-polarized current
is induced by static magnetic fields applied on the three arms of the
$T$-shape QD-array. We calculate the transmission probability in terms of the
scattering matrix which is spin-polarization dependent in the presence of
magnetic fields and leads to the spin-polarized currents. It is the main goal
of the present study to show how the spin currents in output terminals can be
generated and controlled. Particularly two spin currents of
opposite-spin-polarization flowing simultaneously out to electrodes are
obtained and seem extremely useful in the design of quantum- logic-gate devices.

\section{\textbf{MODEL AND FORMULATION}}

The system consisting of three perfect wires (electrodes) attached to a
$T$-shape QD-array of arbitrary number of QDs is explained schematically in
Fig.1. The horizontal QD-array is denoted by the lattice sites $(-N_{L},0),$
$(-N_{L}+1,0),$ $...,(0,0),...,(N_{R}-1,0),(N_{R},0)$ and the vertical
QD-array by the lattice sites $(0,1),$ $(0,2),$ $...,(0,N_{U})$ where $N_{L}$,
$N_{R}$, and $N_{U}$ are the numbers of QDs on the left, right and vertical
arms respectively. Three perfect wires are also described in terms of discrete
lattice sites. The first magnetic field in the z-direction is assumed to be
applied ,for example, on site $(-N_{L},0)$, the second magnetic field on the
site $(N_{R},0)$ is along a direction of angle $\theta_{1}$ with the z-axis,
while the third one on the site $(0,N_{U})$ has a direction angle $\theta_{2}%
$. The latter two magnetic fields not in the same direction as the first one
result in the spin-flip of electrons, which is the mechanism generating the
spin-polarized transport in the QD-array (see the Hamiltonian below). We
remark that it is not necessary to have the three fields to apply on the
certain single QD and what we choose the three QDs is only for the convenience
of analysis. It may be more easy in practical experiments to superpose the
external fields on the three arms of the $T$-shape QD-array. The Hamiltonian
of the system includes five parts
\begin{equation}
H=H_{d,h}+H_{d,v}+H_{el,h}+H_{el,v}+H_{d-el}, \tag{1}%
\end{equation}
where
\begin{align*}
H_{d,h}  &  =%
{\displaystyle\sum\limits_{n=-N_{L}+1,\sigma}^{N_{R}-1}}
\varepsilon_{_{n,0}}d_{_{(n,0),\sigma}}^{+}d_{_{(n,0),\sigma}}+(\varepsilon
_{_{-N_{L},0}}+\sigma g\mu_{_{B}}B)d_{_{(-N_{L},0),\sigma}}^{+}d_{(_{-N_{L}%
,0),\sigma}}\\
&  +(\varepsilon_{N_{R},0}+\sigma g\mu_{_{B_{1}}}B_{1}\cos\theta
_{1})d_{_{(N_{R},0),\sigma}}^{+}d_{_{(N_{R},0),\sigma}}+g\mu_{_{B_{1}}}%
B_{1}\sin\theta_{1}(d_{_{(N_{R},0),\sigma}}^{+}d_{(_{N_{R},0),-\sigma}}+h.c)\\
&  -%
{\displaystyle\sum\limits_{n=-N_{L}+1,\sigma}^{N_{R}-1}}
(td_{_{(n+1,0),\sigma}}^{+}d_{_{(n,0),\sigma}}+h.c)-(t^{^{\prime}}%
d_{_{(N_{R}-1,0),\sigma}}^{+}d_{_{(N_{R},0),\sigma}}+h.c)-(t^{^{\prime}%
}d_{_{(-N_{L}+1,0),\sigma}}^{+}d_{_{(-N_{L},0),\sigma}}+h.c)
\end{align*}
and
\begin{align*}
H_{d,v}  &  =%
{\displaystyle\sum\limits_{n=0,\sigma}^{N_{U}-1}}
\varepsilon_{_{0,n}}d_{_{(0,n),\sigma}}^{+}d_{_{(0,n),\sigma}}\\
&  +(\varepsilon_{0,N_{U}}+\sigma g\mu_{_{B_{2}}}B_{2}\cos\theta
_{2})d_{_{(0,N_{u}),\sigma}}^{+}d_{(_{0,N_{u}),\sigma}}+g\mu_{_{B_{2}}}%
B_{2}\sin\theta_{2}(d_{_{(0,N_{u}),\sigma}}^{+}d_{(_{0,N_{u}),-\sigma}}+h.c)\\
&  -%
{\displaystyle\sum\limits_{n=0,\sigma}^{N_{U}-1}}
(td_{_{(0,n+1),\sigma}}^{+}d_{_{(0,n),\sigma}}+h.c)-(t^{^{\prime}%
}d_{_{(0,N_{U}),\sigma}}^{+}d_{_{(0,N_{U}-1),\sigma}}+h.c)
\end{align*}
denote the Hamiltonians of the horizontal and vertical QD-arrays respectively
with $\varepsilon_{n,0}$ and $\varepsilon_{0,n}$ being the energy eigenvalues
of corresponding QDs. $d_{_{(n,0),\sigma}}^{+}$and $d_{_{(0,n),\sigma}}^{+}$
(with $\sigma=\uparrow,\downarrow$) are the creation operators of the
electrons with spin index $\sigma$ on the QDs, where $-\sigma$ denotes the
opposite spin polarization with respect to $\sigma$. The Zeeman terms induced
by the external fields on the sites $(N_{R},0)$, and $(0,N_{U})$ result in the
spin-flip. It should be emphasized that the $\theta_{1}$ and $\theta_{2}$
dependent terms describing the spin-flip in the Hamiltonian are the key
mechanism of the spin-polarized transport in our system. The matrix elements
defined by $\left\langle n+1,0\right|  H\left|  n,0\right\rangle =\left\langle
0,n+1\right|  H\left|  0,n\right\rangle =t$ denote the hopping integrals
between the nearest neighbors of QDs and are independent of spin polarization
, except for $n=N_{R} $ , $n=-N_{L}-1$, and $n=N_{U}$, where Zeeman energy
induced by the external magnetic fields leads to the energy-level splitting
and therefore the imbalance of tunnel couplings between spin-up and spin-down
electrons. The hopping matrix elements in connection with the QDs $(N_{R},0)$,
$(-N_{L},0)$ and $(0,N_{U})$ are assumed to be $\left\langle N_{R},0\right|
H\left|  N_{R}-1,0\right\rangle =\left\langle -N_{L},0\right|  H\left|
-N_{L}+1,0\right\rangle =\left\langle 0,N_{U}\right|  H\left|  0,N_{U}%
-1\right\rangle =t^{^{\prime}}$ . The Hamiltonians of horizontal and vertical
electrodes are written as
\begin{align*}
H_{el,h}  &  =%
{\displaystyle\sum\limits_{n\leq-(N_{L}+1),\sigma}}
\varepsilon_{_{L}}a_{_{(n,0),\sigma}}^{+}a_{_{(n,0),\sigma}}-%
{\displaystyle\sum\limits_{n\leq-(N_{L}+2),\sigma}}
(ta_{_{(n+1,0),\sigma}}^{+}a_{_{(n,0),\sigma}}+h.c)\\
&  +%
{\displaystyle\sum\limits_{n\geq N_{R}+1,\sigma}}
\varepsilon_{_{R}}a_{_{(n,0),\sigma}}^{+}a_{_{(n,0),\sigma}}-%
{\displaystyle\sum\limits_{n\geq N_{R}+2,\sigma}}
(ta_{_{(n+1,0),\sigma}}^{+}a_{_{(n,0),\sigma}}+h.c),
\end{align*}
\[
H_{el,v}=%
{\displaystyle\sum\limits_{n\geq N_{U}+1,\sigma}}
\varepsilon_{_{U}}a_{_{(0,n),\sigma}}^{+}a_{_{(0,n),\sigma}}-%
{\displaystyle\sum\limits_{n\geq N_{U}+2,\sigma}}
(ta_{_{(0,n+1),\sigma}}^{+}a_{_{(0,n),\sigma}}+h.c),
\]
where $\varepsilon_{_{R}}$ , $\varepsilon_{_{L}}$ and $\varepsilon_{_{U}}$ are
the on-site energies in the right-hand-side, the left-hand-side \ and the
vertical electrodes respectively. $a_{_{(n,0),\sigma}}^{+}$ and
$a_{_{(0,n),\sigma}}^{+}$ are the creation operators of electrons in the
electrodes. Finally the hopping of electrons between the QD and electrode is
described by
\[
H_{d-el}=-V_{_{_{L,\sigma}}}(a_{_{(-N_{L}-1,0)},\sigma}^{+}d_{_{(-N_{L}%
,0),\sigma}}+h.c)-V_{_{_{R,\sigma}}}(a_{_{(N_{R}+1,0),\sigma}}^{+}%
d_{_{(N_{R},0),\sigma}}+h.c)-V_{_{_{U,\sigma}}}(a_{_{(0,N_{U}+1),\sigma}}%
^{+}d_{_{(0,N_{U}),\sigma}}+h.c),
\]
where $V_{R,\sigma}$, $V_{L,\sigma}$, $V_{U,\sigma}$ respectively denote the
tunnel couplings between the three special QDs, where the magnetic fields are
applied, and corresponding electrodes.

The main quantities which we have to calculate in our formulation are the
transmission probabilities $T_{ij}$ of electrons from electrode-$j$ to
electrode-$i$, and the reflection probabilities $R_{jj}$ in the electrode-$j$
which is considered as input terminal. For our three-terminal system we have
$j=1$ (the left electrode) and $i=2,3$ (the right and vertical electrodes)
with $T_{21}$ denoting the transmission probability from the left to right
electrodes and $T_{31}$ from the left to vertical electrodes. However the
situation is not simple for the evaluation of transmission and reflection
probabilities in the multi-terminal system since the electron transport
between two electrodes is by no means an isolated process but affected by
other electrodes. To this end we begin with the stationary Shr$\ddot{o}$dinger
equation
\begin{equation}
\ H\left|  \Psi\right\rangle =E\left|  \Psi\right\rangle , \tag{2}%
\end{equation}
which possesses a general solution of the form
\begin{equation}
\ \left|  \Psi\right\rangle =%
{\displaystyle\sum\limits_{n=-\infty,\sigma}^{\infty}}
C_{(n,0),\sigma}a_{(n,0),\sigma}^{+}\left|  0\right\rangle +%
{\displaystyle\sum\limits_{n=1,\sigma}^{\infty}}
C_{(0,n),\sigma}a_{(0,n),\sigma}^{+}\left|  0\right\rangle . \tag{3}%
\end{equation}
Inserting Eq.(3) into the Shr$\ddot{o}$dinger equation (2) yields
\begin{align}
(E-\varepsilon_{n,0})C_{(n,0),\sigma}+tC_{(n-1,0),\sigma}+tC_{(n+1,0),\sigma}
&  =0,\ n\neq-N_{L},\text{ }0,N_{R},\nonumber\\
(E-\varepsilon_{0,n})C_{(0,n),\sigma}+tC_{(0,n-1),\sigma}+tC_{(0,n+1),\sigma}
&  =0,\ n\neq\text{ }N_{U},(n\geq1)\nonumber\\
(E-\varepsilon_{0,0})C_{(0,0),\sigma}+tC_{(0,1),\sigma}+tC_{(-1,0),\sigma
}+tC_{(1,0),\sigma}  &  =0,\text{ }n=0 \tag{4}%
\end{align}
\begin{equation}
(E-(\varepsilon_{-N_{L},0}+\sigma g\mu_{B}B))C_{(-N_{L},0),\sigma}%
+V_{L,\sigma}C_{-N_{L}-1,\sigma}+t^{\prime}C_{-N_{L}+1,\sigma}=0, \tag{5}%
\end{equation}
\begin{equation}%
\begin{tabular}
[c]{l}%
$(E-(\varepsilon_{N_{R},0}+\sigma g\mu_{B_{1}}B_{1}\cos\theta_{1}%
))C_{(N_{R},0),\sigma}+B_{1}\sin\theta_{1}C_{(N_{R},0),-\sigma}$\\
$+t^{\prime}C_{(N_{R}-1,0),\sigma}+V_{R,\sigma}C_{(N_{R}+1,0),\sigma}=0,$%
\end{tabular}
\tag{6}%
\end{equation}
\begin{equation}%
\begin{tabular}
[c]{l}%
$(E-(\varepsilon_{0,N_{U}}+\sigma g\mu_{B_{2}}B_{2}\cos\theta_{2}%
))C_{(0,N_{U}),\sigma}+B_{2}\sin\theta_{2}C_{(0,N_{U}),-\sigma}$\\
$+t^{\prime}C_{(0,N_{U}-1),\sigma}+V_{U,\sigma}C_{(0,N_{U}+1),\sigma}=0.$%
\end{tabular}
\tag{7}%
\end{equation}
Notice that in the input electrode-$1$ the wave function is a superposition of
incoming plane wave of unit amplitude and a reflection wave with amplitude
$r_{11}(\sigma)$, while in the output electrodes-$2$, $3$ the outgoing plane
waves possess transmission amplitudes $t_{21}(\sigma)$ and $t_{31}(\sigma)$
respectively, we have
\begin{equation}
\ \ C_{(n,0),\sigma}=e^{iK_{\sigma}^{L}na}+r_{11}(\sigma)e^{-iK_{\sigma}%
^{L}na},\ \ \ \ \ \ n\leq-(N_{L}+1), \tag{10}%
\end{equation}
\begin{equation}
C_{(n,0),\sigma}=t_{21}(\sigma)e^{iK_{\sigma}^{R}na},\text{\ \ \ }n\geq
N_{R}+1, \tag{11}%
\end{equation}
\begin{equation}
C_{(0,n),\sigma}=t_{31}(\sigma)e^{iK_{\sigma}^{U}na},\text{\ \ \ }n\geq
N_{U}+1, \tag{12}%
\end{equation}
where, $K_{\sigma}^{L}$, $K_{\sigma}^{R}$, and $K_{\sigma}^{U}$ are the wave
vectors in the left-, right-, and vertical-electrode respectively. Using these
wave functions, we can eliminate all the coefficients, $\{C_{(0,n),\sigma}\}$,
in Eg.$(4)$ and Eg.$(7)$ and obtain
\[
(E-\varepsilon_{n,0})C_{(n,0),\sigma}+tC_{(n-1,0),\sigma}+tC_{(n+1,0),\sigma
}=0,\ n\neq-N_{L},\text{ }0,N_{R},
\]
\[
(E-(\varepsilon_{0,0}+\varepsilon_{0,0}^{U})C_{(0,0),\sigma}+tC_{(0,1),\sigma
}+tC_{(-1,0),\sigma}+tC_{(1,0),\sigma}=0,n=0
\]
\[
(E-(\varepsilon_{-N_{L},0}+\sigma g\mu_{B}B))C_{(-N_{L},0),\sigma}%
+V_{L,\sigma}C_{-N_{L}-1,\sigma}+t^{\prime}C_{-N_{L}+1,\sigma}=0,
\]
\begin{equation}%
\begin{tabular}
[c]{l}%
$(E-(\varepsilon_{N_{R},0}+\sigma g\mu_{B_{1}}B_{1}\cos\theta_{1}%
))C_{(N_{R},0),\sigma}+B_{1}\sin\theta_{1}C_{(N_{R},0),-\sigma}$\\
$+t^{\prime}C_{(N_{R}-1,0),\sigma}+V_{R,\sigma}C_{(N_{R}+1,0),\sigma}=0,$%
\end{tabular}
\tag{13}%
\end{equation}
where $\varepsilon_{0,0}+\varepsilon_{0,0}^{U}$ is the effective energy on
site $(0,0)$ with
\begin{equation}
\varepsilon_{0,0}^{U}=\frac{V_{U,\sigma}^{2}}{E-\varepsilon_{0,1}-\frac{t^{2}%
}{E-\varepsilon_{0,2}-\frac{t^{2}}{%
\begin{array}
[c]{c}%
.\\
\text{ \ \ \ \ \ \ \ \ }.\\
\text{ \ \ \ \ \ \ \ \ \ \ \ \ \ \ \ \ \ \ }.\\
\text{ \ \ \ \ \ \ \ \ \ \ \ \ }\frac{(t^{\prime})^{2}}{E-\varepsilon
_{0,N_{U}}-{\sum\nolimits^{U}}(E)}.
\end{array}
}}} \tag{14}%
\end{equation}
Here $%
{\displaystyle\sum\nolimits^{U}}
(E)=V_{U,\sigma}^{2}G^{U}(E)$ denotes the self-energy due to the coupling with
the horizontal QD-array, and $G^{U}(E)$ is the Green function of the
lattice-site $(0,N_{U})$ satisfying the recursive equation
\begin{equation}
G^{U}(E)=[E-\varepsilon_{U}-t^{2}G^{U}(E)]^{-1}, \tag{15}%
\end{equation}
the solution of which is seen to be
\begin{equation}
G^{U}(E)=\frac1{2t^{2}}\{(E-\varepsilon_{U})-i[4t^{2}-(E-\varepsilon_{U}%
)^{2}]^{\frac12}\}. \tag{16}%
\end{equation}

We can see that, the Eq. (13) does not include any of the expansion
coefficients, $\{C_{(0,n),\sigma}\}$, associated with the vertical
lattice-site. Thus it implies that the three-terminal device reduces to an
effective two-terminal system and the effect of the vertical QD-array appears
as a self-energy $%
{\displaystyle\sum\nolimits^{U}}
(E)$ which has now been included in the effective on-site energy.

Now we rewrite Eq. (13) as a matrix equation
\[
\ \ \ \ \ \ \ \ \ \left[
\begin{array}
[c]{c}%
C_{(n+1,0),\sigma}\\
C_{(n,0),\sigma}\\
C_{(n+1,0),-\sigma}\\
C_{(n,0),-\sigma}%
\end{array}
\right]  =M[(n,0),E]\left[
\begin{array}
[c]{c}%
C_{(n,0),\sigma}\\
C_{(n-1,0),\sigma}\\
C_{(n,0),-\sigma}\\
C_{(n-1,0),-\sigma}%
\end{array}
\right]  ,
\]
where $M[(n,0),E]$ is called the transfer matrix which links the expansion
coefficient vector $(C_{(n+1,0),\sigma},C_{(n,0),\sigma},C_{(n+1,0),-\sigma
},C_{(n,0),-\sigma})^{T}$ to the vector \ $(C_{(n,0),\sigma},C_{(n-1,0),\sigma
},C_{(n,0),-\sigma},C_{(n-1,0),-\sigma})^{T}$ and thus is defined by
\begin{align}
\ \ M[(n,0),E]  &  =\left[
\begin{array}
[c]{cccc}%
-\frac{E-\varepsilon_{n,0}^{\prime}}t & -1 & 0 & 0\\
1 & 0 & 0 & 0\\
0 & 0 & -\frac{E-\varepsilon_{n,0}^{\prime}}t & -1\\
0 & 0 & 1 & 0
\end{array}
\right]  =\left[
\begin{array}
[c]{cc}%
Q_{1} & 0\\
0 & Q_{2}%
\end{array}
\right]  ,\ n\neq N_{R}\ \nonumber\\
M[(N_{R},0),E]  &  =\left[
\begin{array}
[c]{cccc}%
-\frac{E-\varepsilon_{N_{R},0}^{\prime}}{V_{R,\sigma}} & -\frac{t^{\prime}%
}{V_{R,\sigma}} & \frac{g\mu_{B_{1}}B_{1}\sin\theta_{1}}{V_{R,\sigma}} & 0\\
1 & 0 & 0 & 0\\
\frac{g\mu_{B_{1}}B_{1}\sin\theta_{1}}{V_{R,\sigma}} & 0 & -\frac
{E-\varepsilon_{N_{R},0}^{\prime}}{V_{R,\sigma}} & -\frac{t^{\prime}%
}{V_{R,\sigma}}\\
0 & 0 & 1 & 0
\end{array}
\right]  ,\ \ (n=N_{R}). \tag{17}%
\end{align}
where $\varepsilon_{(n,0)}^{\prime}$ is the effective on-site energy given by
\begin{align}
\varepsilon_{n,0}^{\prime}  &  =\varepsilon_{n,0},\text{ }n\neq-N_{L}%
,0,N_{R}\nonumber\\
\varepsilon_{-N_{L},0}^{\prime}  &  =\varepsilon_{-N_{L},0}+\sigma g\mu
_{B}B,\text{ }n=-N_{L}\nonumber\\
\varepsilon_{0,0}^{\prime}  &  =\varepsilon_{0,0}+\varepsilon_{0,0}^{U},\text{
}n=0,\nonumber\\
\varepsilon_{N_{R},0}^{\prime}  &  =\varepsilon_{N_{R},0}+\sigma g\mu_{B_{1}%
}B_{1}\cos\theta_{1},\text{ }n=-N_{R}. \tag{18}%
\end{align}
From Eqs. $(10)$ and $(11)$, it can be shown that the transmission amplitude
$t_{21}(\sigma)$ is related to the incident coefficients via the equation
\begin{equation}
\left[
\begin{array}
[c]{c}%
t_{21}(\sigma)\\
0\\
t_{21}(-\sigma)\\
0
\end{array}
\right]  =T(E)\left[
\begin{array}
[c]{c}%
1\\
r_{11}(\sigma)\\
1\\
r_{11}(-\sigma)
\end{array}
\right]  , \tag{19}%
\end{equation}
where the transfer matrix $T(E)$ for the whole system is given by
\begin{align}
T(E)  &  =\left[
\begin{array}
[c]{cccc}%
e^{-iK_{\sigma}^{R}(N_{R}+1)a} & 0 & 0 & 0\\
0 & e^{iK_{\sigma}^{R}(N_{R}+1)a} & 0 & 0\\
0 & 0 & e^{-iK_{-\sigma}^{R}(N_{R}+1)a} & 0\\
0 & 0 & 0 & e^{iK_{-\sigma}^{R}(N_{R}+1)a}%
\end{array}
\right]  \times\nonumber\\
&  \left[
\begin{array}
[c]{cccc}%
e^{ik_{\sigma}^{R}a} & e^{-ik_{\sigma}^{R}a} & 0 & 0\\
1 & 1 & 0 & 0\\
0 & 0 & e^{ik_{-\sigma}^{R}a} & e^{-ik_{-\sigma}^{R}a}\\
0 & 0 & 1 & 1
\end{array}
\right]  ^{-1}%
{\displaystyle\prod\limits_{n=N_{R}+1}^{-(N_{L}+1)}}
M[(n,0),E]\nonumber\\
&  \times\left[
\begin{array}
[c]{cccc}%
e^{ik_{\sigma}^{L}a} & e^{-ik_{\sigma}^{L}a} & 0 & 0\\
1 & 1 & 0 & 0\\
0 & 0 & e^{ik_{-\sigma}^{L}a} & e^{-ik_{-\sigma}^{L}a}\\
0 & 0 & 1 & 1
\end{array}
\right]  \left[
\begin{array}
[c]{cccc}%
e^{-iK_{\sigma}^{L}(N_{L}+1)a} & 0 & 0 & 0\\
0 & e^{iK_{\sigma}^{L}(N_{L}+1)a} & 0 & 0\\
0 & 0 & e^{-iK_{-\sigma}^{L}(N_{L}+1)a} & 0\\
0 & 0 & 0 & e^{iK_{-\sigma}^{L}(N_{L}+1)a}%
\end{array}
\right]  . \tag{20}%
\end{align}
Thus from Eq.$(19)$ we can obtain the reflection amplitudes $r_{11}(\uparrow)$
and $r_{11}(\downarrow)$ as%

\begin{equation}
r_{11}(\uparrow)=\frac{-(\frac{T_{21}}{T_{22}}+\frac{T_{23}}{T_{22}}%
)+\frac{T_{24}}{T_{22}}(\frac{T_{41}}{T_{44}}+\frac{T_{43}}{T_{44}})}%
{1-\frac{T_{24}}{T_{22}}\frac{T_{42}}{T_{44}}}, \tag{21}%
\end{equation}

\begin{equation}
r_{11}(\downarrow)=\frac{-(\frac{T_{41}}{T_{44}}+\frac{T_{43}}{T_{44}}%
)+\frac{T_{42}}{T_{44}}(\frac{T_{21}}{T_{22}}+\frac{T_{23}}{T_{22}})}%
{1-\frac{T_{24}}{T_{22}}\frac{T_{42}}{T_{44}}}, \tag{22}%
\end{equation}
and the transmission amplitudes $t_{21}(\uparrow)$ and $t_{21}(\downarrow)$%
\begin{equation}
t_{21}(\uparrow)=T_{11}+T_{12}r_{11}(\uparrow)+T_{13}+T_{14}r_{11}%
(\downarrow), \tag{23}%
\end{equation}
\begin{equation}
t_{21}(\downarrow)=T_{31}+T_{32}r_{11}(\uparrow)+T_{33}+T_{34}r_{11}%
(\downarrow). \tag{24}%
\end{equation}
The spin-polarized reflection and transmission probabilities are seen to be
\begin{equation}
R_{11}(\sigma)=\frac12\left|  r_{11}(\sigma)\right|  ^{2}, \tag{25}%
\end{equation}
\begin{equation}
T_{21}(\sigma)=\frac12\frac{V_{R,\sigma}}{V_{L,\sigma}}\left|  t_{21}%
(\sigma)\right|  ^{2}, \tag{26}%
\end{equation}
and the total transmission and reflection probabilities are defined by
\begin{equation}
T_{21,tot}=T_{21}(\uparrow)+T_{21}(\downarrow), \tag{27}%
\end{equation}
\begin{equation}
R_{11,tot}=R_{21}(\uparrow)+R_{21}(\downarrow). \tag{28}%
\end{equation}
The spin current can be derived from the difference between the two
spin-polarized transmission probabilities
\begin{equation}
\Delta T_{21}=T_{21}(\uparrow)-T_{21}(\downarrow). \tag{29}%
\end{equation}
The transmission probability, $T_{31}(\sigma)$, from electrode-$1$ to
electrode-$3$ can be simply obtained by replacing the on-site energy
$\varepsilon_{0,0}^{\prime}=\varepsilon_{0,0}+\varepsilon_{0,0}^{U}$ in the
matrix $M[(0,0),E]$ with $\varepsilon_{0,0}^{\prime}=\varepsilon
_{0,0}+\varepsilon_{0,0}^{R}$ where
\begin{equation}
\varepsilon_{0,0}^{R}=\frac{V_{R,\sigma}^{2}}{E-\varepsilon_{1,0}-\frac{t^{2}%
}{E-\varepsilon_{2,0}-\frac{t^{2}}{%
\begin{array}
[c]{c}%
.\\
\text{ \ \ \ \ \ \ \ \ }.\\
\text{ \ \ \ \ \ \ \ \ \ \ \ \ \ \ \ \ \ \ }.\\
\text{ \ \ \ \ \ \ \ \ \ \ \ \ }\frac{(t^{\prime})^{2}}{E-\varepsilon
_{N_{R},0}-{\sum\nolimits^{R}}(E)},
\end{array}
}}} \tag{30}%
\end{equation}
and $%
{\displaystyle\sum\nolimits^{R}}
(E)=V_{R,\sigma}^{2}G^{R}(E)$ is the self-energy due to the presence of the
right electrode, and
\begin{equation}
G^{R}(E)=[E-\varepsilon_{U}-t^{2}G^{R}(E)]^{-1}, \tag{31}%
\end{equation}
is the Green$^{,}$s function of the lattice-site $(N_{R}+1,0)$. Moreover the
effective on-site energy $\varepsilon_{n,0}^{\prime}$ of Eq. $(18)$ should be
replaced by
\begin{align*}
\varepsilon_{0,n}^{\prime}  &  =\varepsilon_{0,n},\text{ }n\neq-N_{L}%
,0,N_{U}\\
\varepsilon_{-N_{L},0}^{\prime}  &  =\varepsilon_{-N_{L},0}+\sigma g\mu
_{B}B,\text{ }n=-N_{L}\\
\varepsilon_{0,0}^{\prime}  &  =\varepsilon_{0,0}+\varepsilon_{0,0}^{R},\text{
}n=0,\\
\varepsilon_{0,N_{U}}^{\prime}  &  =\varepsilon_{0,N_{U}}+\sigma g\mu_{B_{2}%
}B_{2}\cos\theta_{2},\text{ }n=N_{U}.
\end{align*}
Then the transmission probability $T_{31}(\sigma)$ can be evaluated from
\begin{equation}
T_{31}(\sigma)=\frac12\frac{V_{U,\sigma}}{V_{L,\sigma}}\left|  t_{31}%
(\sigma)\right|  ^{2}, \tag{32}%
\end{equation}
and finally we obtain the total and spin transmission probabilities
$T_{31,tot}$, $\Delta T_{31}$ in the same way as for the calculation of
$T_{21,tot}$ and $\Delta T_{21}$.

The conductance of the device can be obtained from the transmission
probability with the help of the Laudauer-B$\ddot{u}$ticker formula
\begin{equation}
G=(\frac{2e^{2}}{h})T, \tag{33}%
\end{equation}
where $e$ is the electron charge, $h$ the Planck's constant, and $T$ the
transmission probability of the device.

\section{\textbf{\ NUMERICAL RESULT AND DISCUSSION}}

\subsection{Homogeneous dot-array}

In the numerical evaluation the lattice constant $a$ is chosen as unit of
length and hopping constant $t$ as the unit of energy. The on-site energies of
all QDs, as well as $\varepsilon_{L}$, $\varepsilon_{R}$ , $\varepsilon_{U}$
are set to the same value of $2t$. The coupling parameters between the QDs and
the three electrodes are assumed to be $V_{L,\sigma}=V_{R,\sigma}=V_{U,\sigma
}=0.5t$ and $g\mu_{B_{1}}B_{1}=g\mu_{B_{2}}B_{2}=1t.$

First of all we consider the simplest case that there is only one-QD on each
arm of the $T$-shape array besides the central dot, i.e. $N_{L}=N_{R}%
=N_{U}=1.$ Figs. $2(a),3(a)$ and $2(b)$, $3(b)$ display the energy-dependence
(spectrum) of the spin-up and the spin-down transmission probabilities to the
electrode-$2$ and the electrode -$3$ respectively (for $t=1$ ,\ $t^{\prime
}=0.8$, $\theta_{1}=\frac34\pi$, $\theta_{2}=\frac14\pi$). The total
transmissions $T_{21,tot}$ , $T_{31,tot}$ are shown in Figs. $4(a)$ and
$4(b)$. It can be seen that in the presence of the external magnetic fields,
the transmitted electrons split into spin-up and spin-down components with
different spectra.\ Fig. $5$ is the plot of the relative transmission
probabilities between two spin components $\Delta T_{21}$( solid line) and
$\Delta T_{31}$ (dotted line) versus the energy respectively. We can see that
\ in some electron-energy ranges, for example from $1.8t$ to $2.4t$ or from
$3.4t$ to $3.8t$, the spin transmission probability to the electrode-$3$
,$\Delta T_{31},$\ is positive (i.e. the spin current with spin-up
polarization) while the spin transmission probability to electrode-$2$,
$\Delta T_{21},$\ is negative (namely the spin-down current). However the
situation would be opposite in other energy region , for example from $0t$ to
$1.2t$ , where $\Delta T_{21}$\ is positive and $\Delta T_{31}$\ is negative.

Figs. $6$ show the spin transition probabilities $\Delta T_{21}$ (solid line)
and $\Delta T_{31}$ (dotted line) as a function of energy for the multi-QD
case that $N_{L}=N_{R}=4$ and $N_{U}=3$ with the magnetic field direction
angles $\theta_{1}=\theta_{2}=\frac{1}{2}\pi$. In the electron-energy range
from $0.6t$ to $0.75t$ or from $1.78t$ to $1.90t$ the spin transmission
probability $\Delta T_{31}$\ is positive (i.e. the spin-up current flowing to
electrode-$3$) while $\Delta T_{21}$\ is negative ( spin-down current flowing
to electrode-$2$). Opposite spin currents are observed in the energy region
from $0.38t$ to $0.47t$ , where $\Delta T_{31}$\ is positive and $\Delta
T_{21}$\ is negative. Fig. $7$ are the results of $\Delta T_{21}$ (solid line)
and $\Delta T_{31}$ (dotted line) for the direction angles $\theta_{1}%
=\frac{3}{4}\pi$ while $\theta_{2}=\frac{1}{4}\pi.$ We see that the spin
currents are sensitive to the fields. It may also be worthy to point out that
in the energy region between $2.9t$ and $3.1t$, $\Delta T_{21}$\ is negligibly
small while $\Delta T_{31}$\ is positive. In other words the spin current can
exist only in one output electrode. Comparing with previous case (one-QD on
each arm) more resonant peaks of the spectra appear with the increasing number
of QDs in the same way as the usual quantum transport in the QD-array without spin-polarization.

The generation of spin currents in the two output electrodes in our model can
be understood that a force applied on the spin-$\mathbf{S}$ which is\ induced
by the spatially inhomogeneous magnetic fields, namely $\mathbf{\ F}=g\mu
_{b}\mathbf{S}\cdot\bigtriangledown\mathbf{B}$ (\ where\textbf{\ }$\mu_{b}$ is
the Bohr magneton, $g$ is the spin g-factor), removes the degeneracy of
spin-polarization in the transport and thus leads to the relative shifts of
resonant peaks of the transmission spectra between the spin-up and spin-down
components of electron.\textbf{\ }The spin current appears when the
transmission spectrum of one spin component reaches a maximum while the
spectrum of the other spin component is a minimum\textbf{.}

\subsection{Effect of disorder}

The homogeneous dot-arry is just a theoretical idealization and disorder is
necessary to be taken into account in practical systems. It is interesting to
see the disorder effect of the QD-array on the spin-polarized \ quantum
transport.\ To this end we consider the disorder following
Refs.\cite{Terraneo,Texie,Pin,Tit,Wei} that the on-site energies of QDs are
alternated with $4(1-x)$\ and $4\cdot x$, where $0\leq x\leq1$\ is the
probability distribution parameter. When $x=0.5$, the system reduces to the
non-disordering one. In the $T$-shape QD-array on-site energies $\varepsilon
_{_{n,0}}$ ($-N_{L}\leq n\leq N_{R}$ , on the horizontal arm) and
$\varepsilon_{_{0,n}}$ ($1\leq n\leq N_{U}$ , on vertical arm) are set to
$4(1-x)$ for odd-$n$ and $4x$ for even-$n$. Fig. (8) is a comparison of plots
of $\Delta T_{31}$\ for the probability distribution parameter $x=0.47$
(dotted line), $0.5$ (solid line) respectively with $N_{L}=N_{R}=4$ and
$N_{U}=3$.\ We can see that the disorder can shift the resonant peaks of the
transmission probability $\Delta T_{31}$ slightly, and suppress the height of
some peaks but no more then 20\% while increase the height of other peaks no
more then 18\%. In other words\ the spin splitting is robust against the disorder.

The latest advances in nanotechnology make it possible to fabricate
QD-array\cite{Orellana} and the model of spin-splitter proposed in this paper
may be realizable experimentally. In practical experiment it is more easy to
apply the external fields on three arms of the $T$-shap QD-array instead of
three QDs. The relative shifts of resonant peaks of the transmission spectra
between the spin-up and spin-down components of electron can be modulated by
the magnetic fields and on-site energies.

\section{SUMMARY AND CONCLUSION}

In this paper we demonstrate theoretically a method to generate and control
spin currents in a three-terminal system. The output spin currents are tunable
in a wide range of magnitudes and various output configurations by adjusting
the energy and the direction angles of the magnetic fields as well. Moreover
it is demonstrated by the numerical evaluation that the spin currents remain
in the presence of the disorder of QD-array. Particularly the spin currents
with opposite spin polarizations in the two output electrodes can be
generated. Thus this device is, as a matter of fact, a spin splitter similar
with the light beam polarimeter in optics. This observation may have practical
application in the fabrication of spintronic devices for logic gates with
spin-up and spin-down outputs regarding as qubits.

\section{ACKNOWLEDGMENTS}

This work was supported by National Natural Science Foundation of China under
Grant Nos. 10475053.

\textbf{FIGURE CAPTIONS}

Fig.(1) $T$-shape QD-array of three-electrode with three magnetic fields
applied on the dots $(-N_{L},0),(N_{R},0)$ and $(0,N_{U})$ respectively.

Fig.(2) Spin-up (a) and spin-down transmission probabilities to the
electrode-$2$ as a function of energy.

\bigskip Fig.(3) Spin-up (a) and spin-down ransmission probabilities to the
electrode-$3$ as a function of energy.

Fig.(4) Total transmission probabilities to electrode -2 and electrode- 3.

Fig.(5) Spin-polarized transmission probabilities to electrode- 2 and electrode-3.

Fig.(6) Spin-polarized\ transmission probabilities to electrode-2 (solid line)
and electrode-3 (dotted line) versus the energy for angles $\theta_{1}%
=\frac{1}{2}\pi$ and $\theta_{2}=\frac{1}{2}\pi$.

Fig.(7) Spin-polarized\ transmission probabilities to electrode-2 (solid line)
and electrode-3 (dotted line) versus the energy for angles $\theta_{1}%
=\frac{3}{4}\pi$ and $\theta_{2}=\frac{1}{4}\pi$.

Fig.(8) The comparision of the plots $\Delta T_{31}$\ with (dotted line) and
without (solid line) the disoder for angles $\theta_{1}=\frac{1}{2}\pi$\ and
$\theta_{2}=\frac{1}{2}\pi$.

\textbf{\qquad\ }

\textbf{\qquad\ }

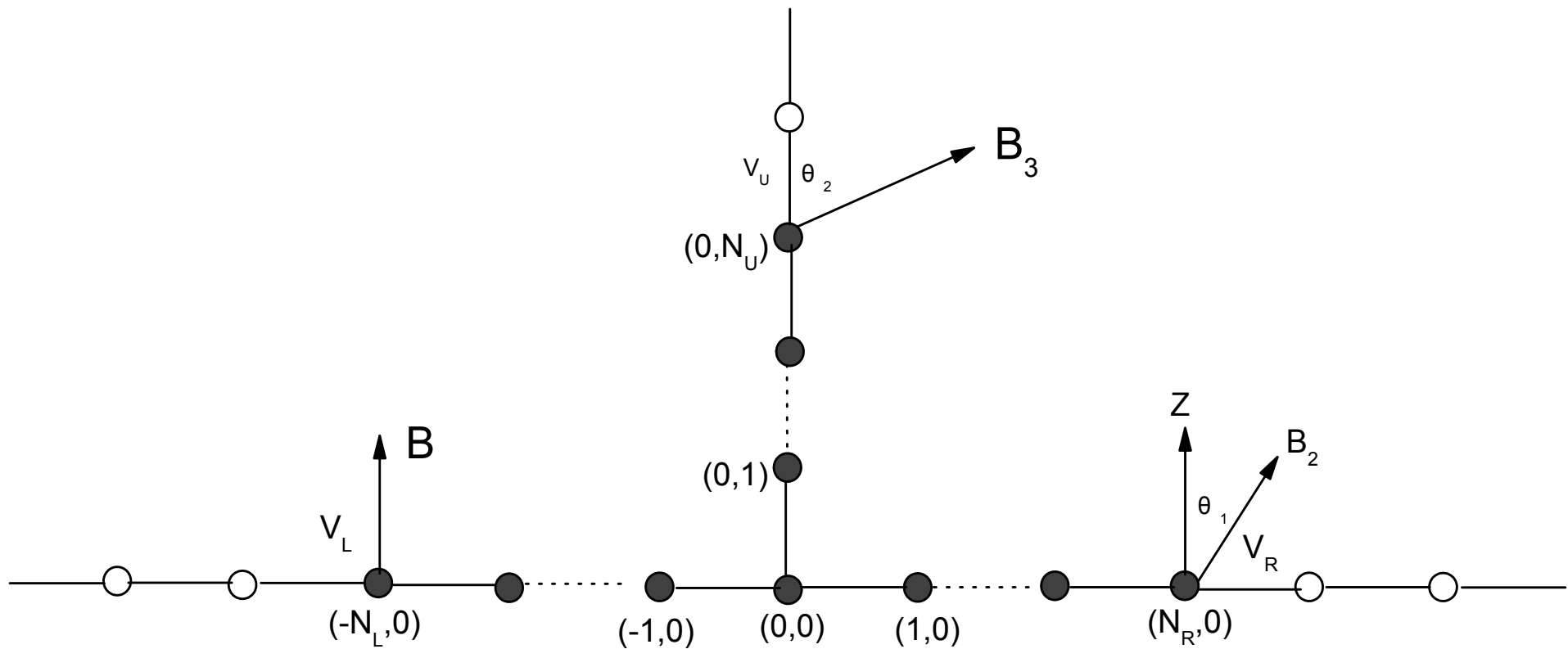
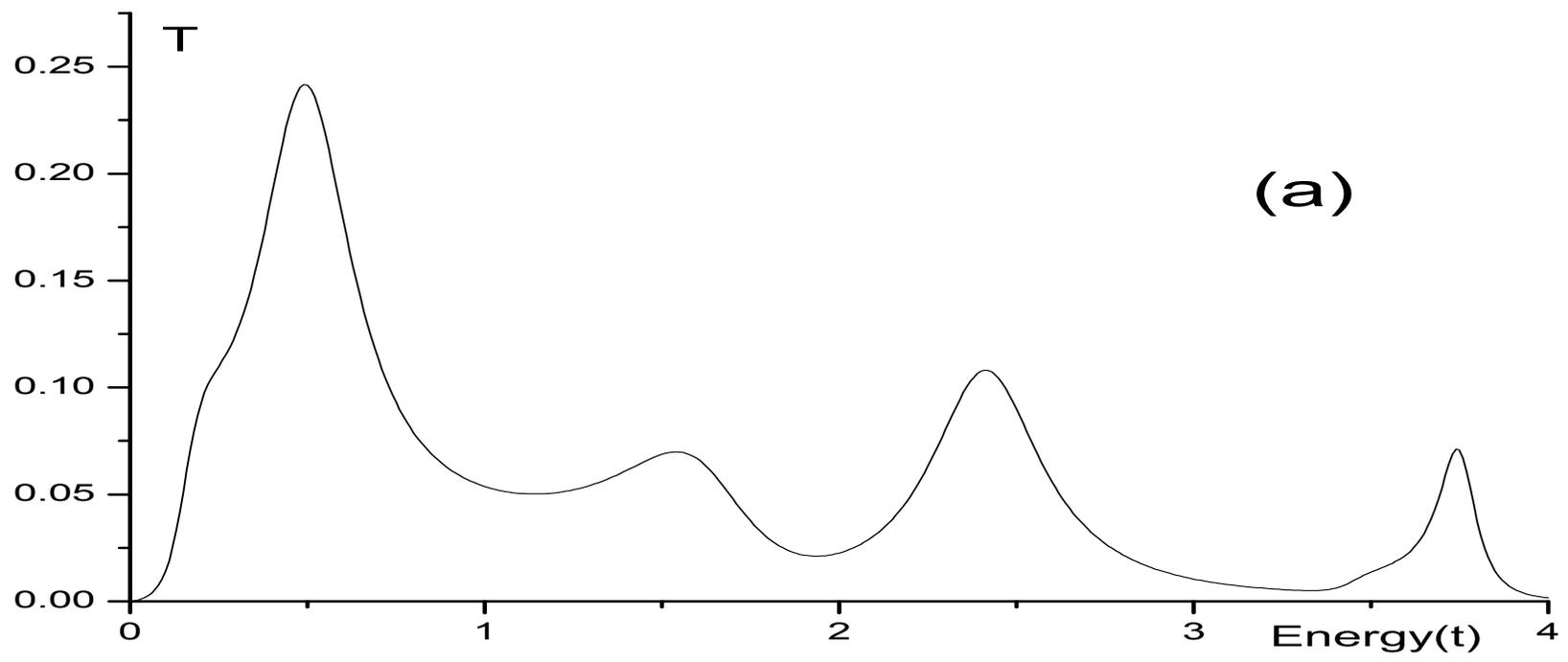
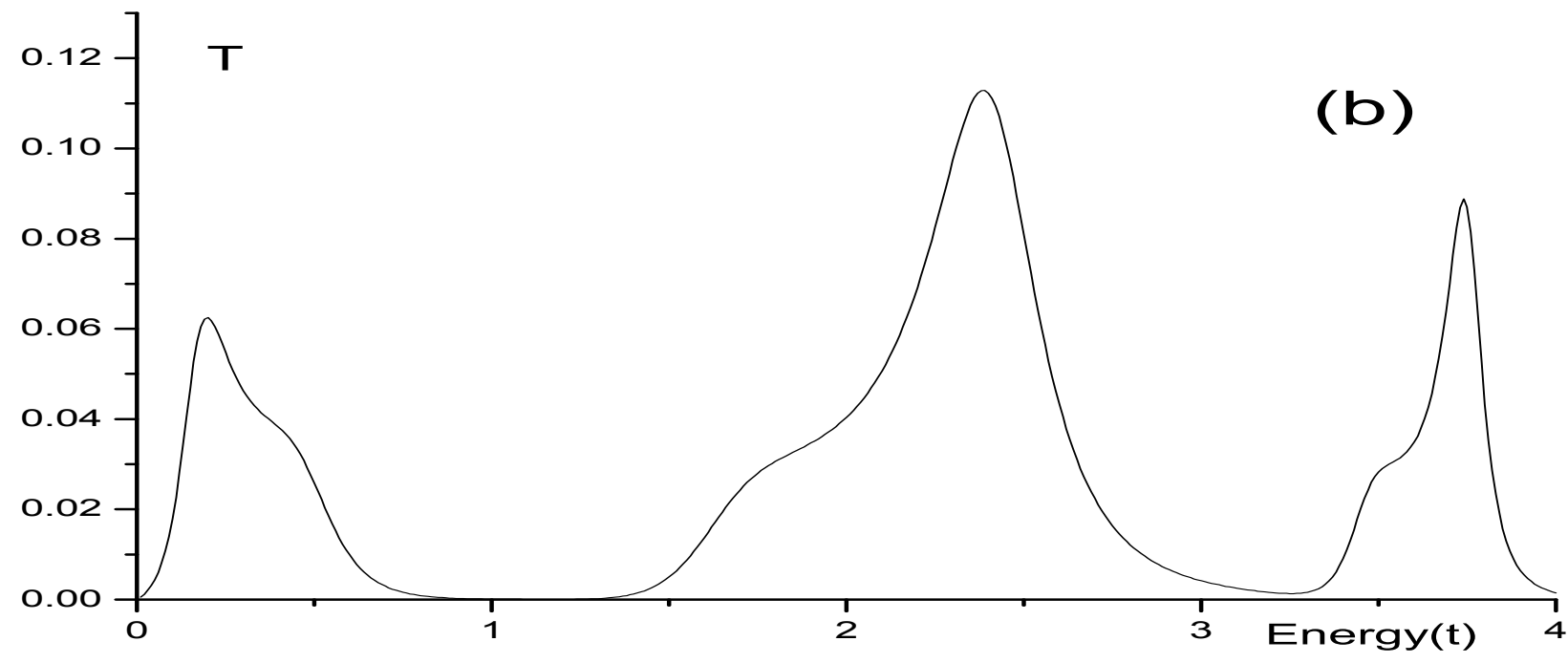
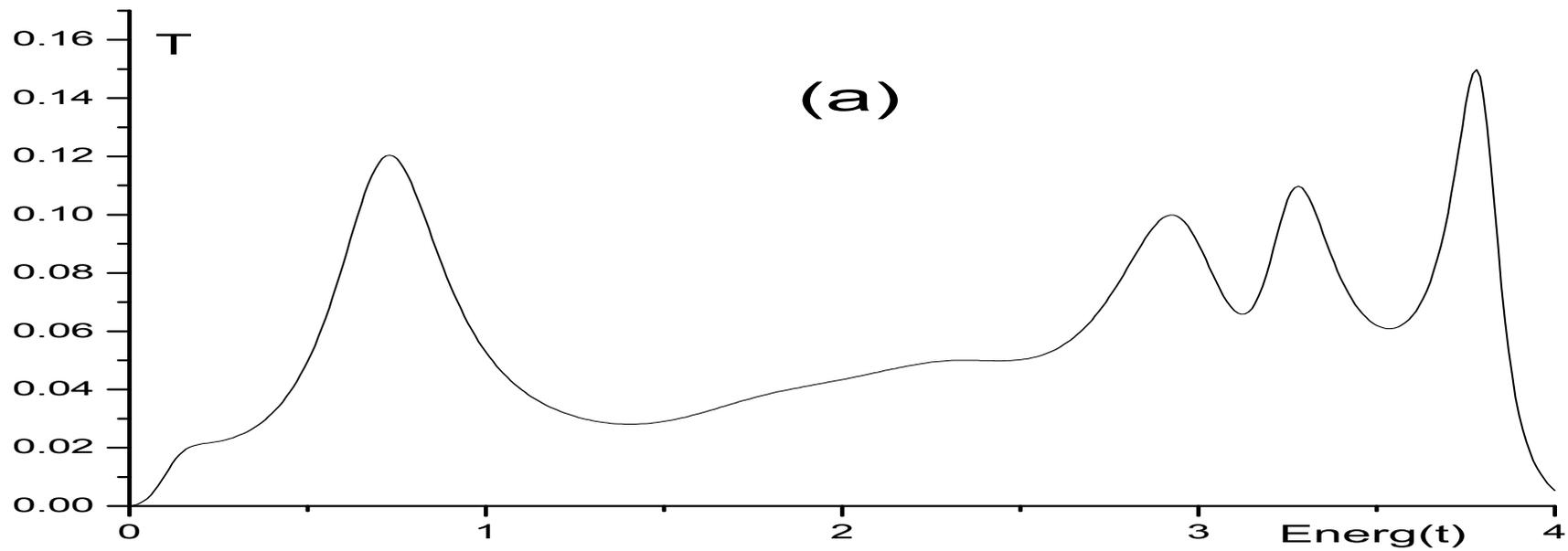
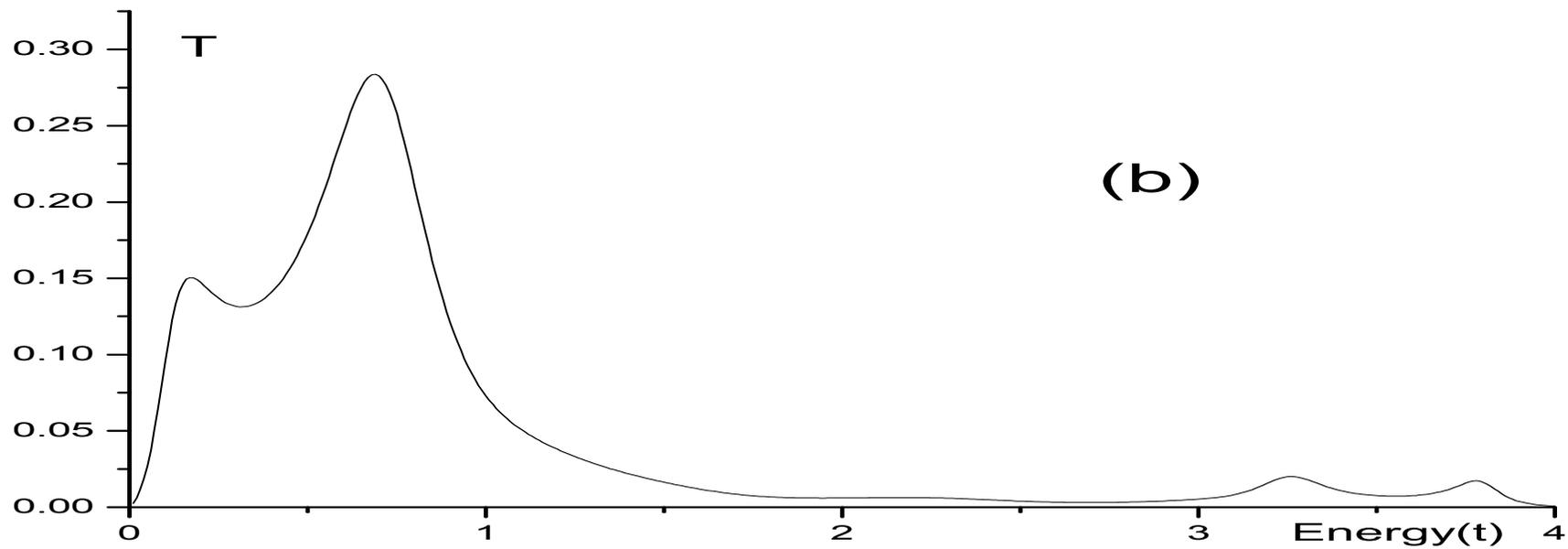
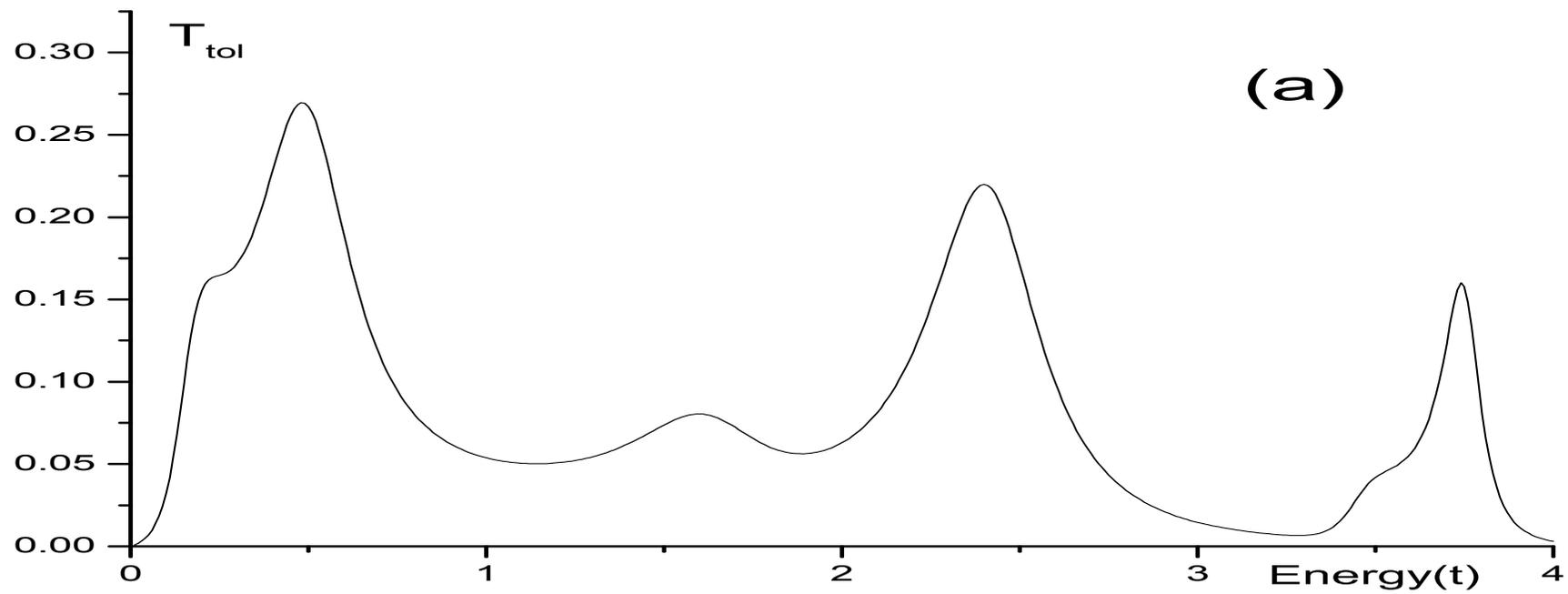
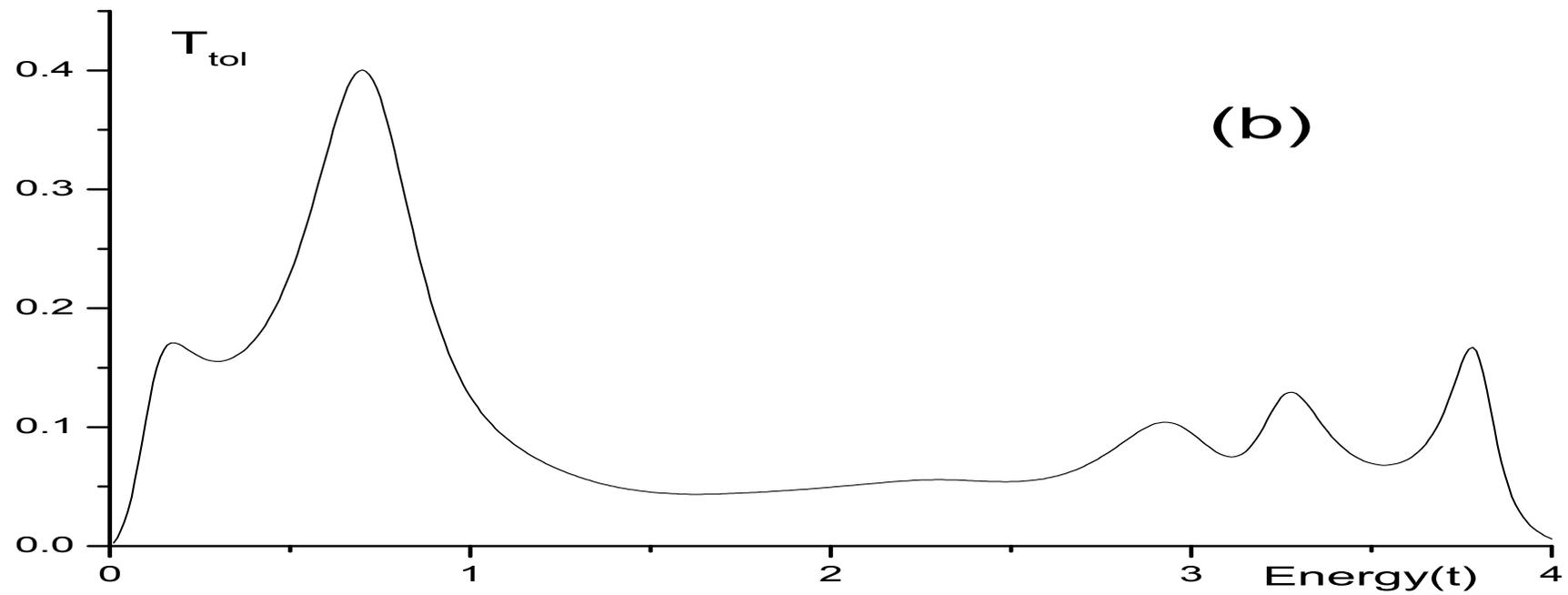
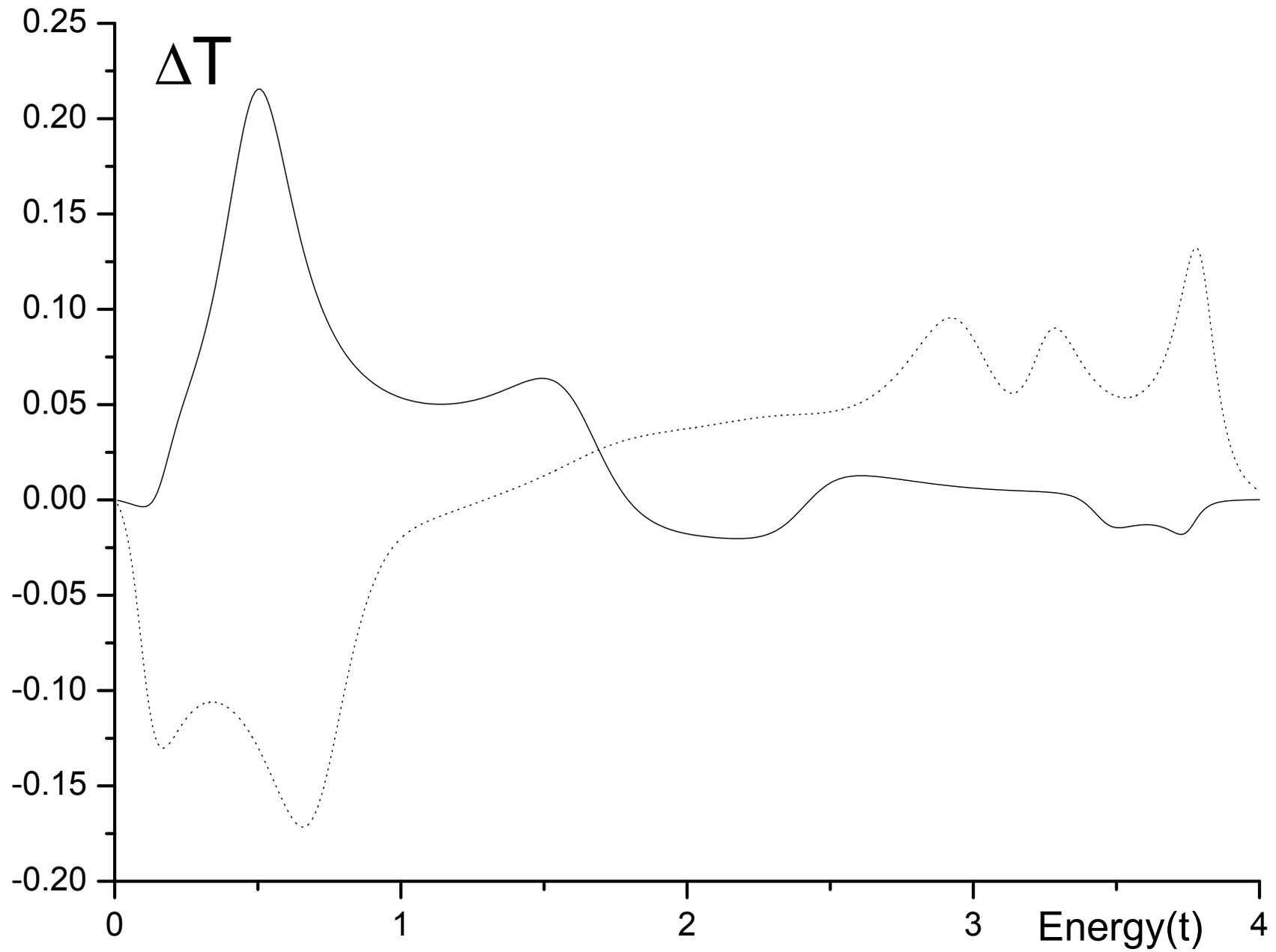
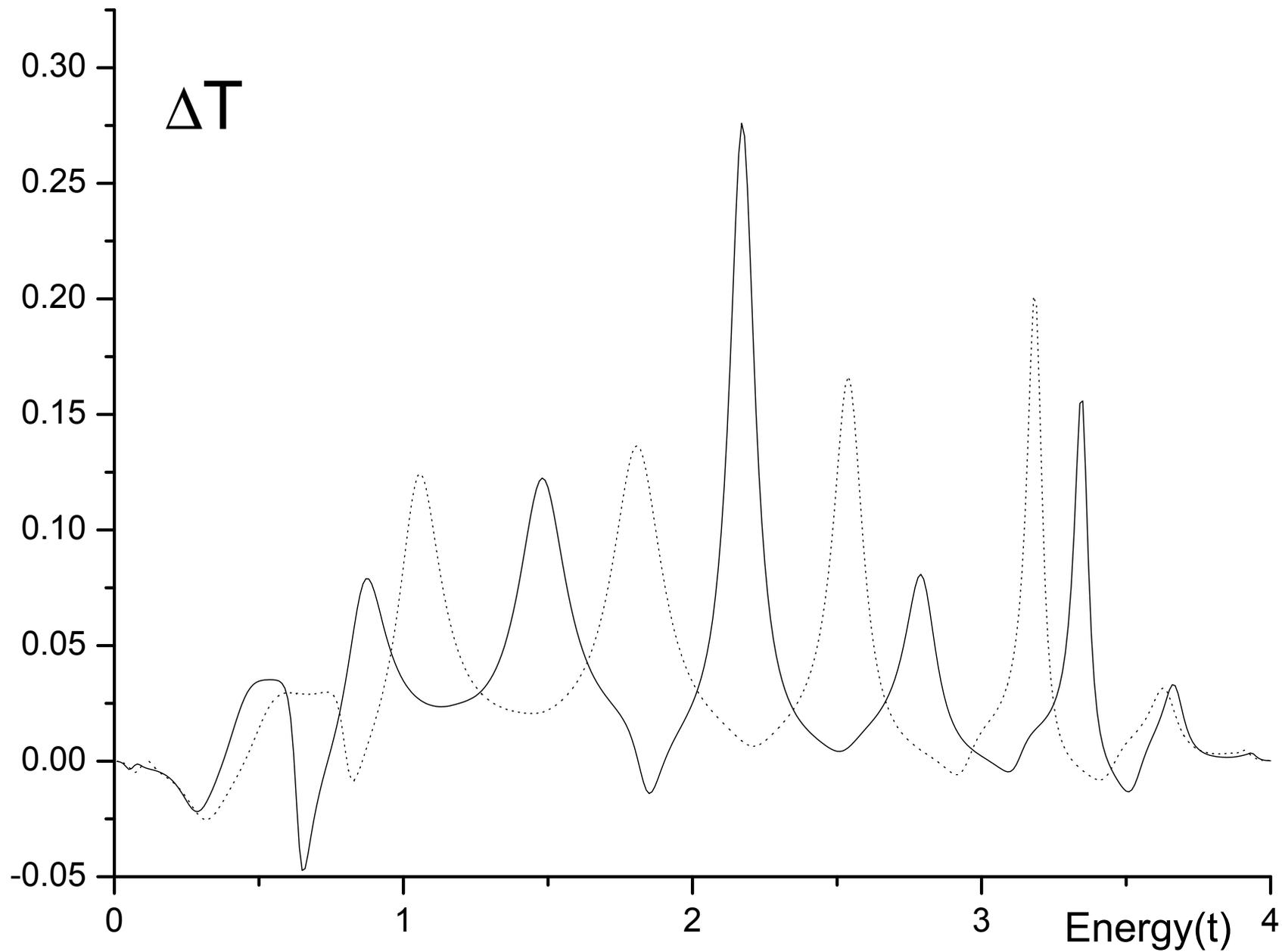
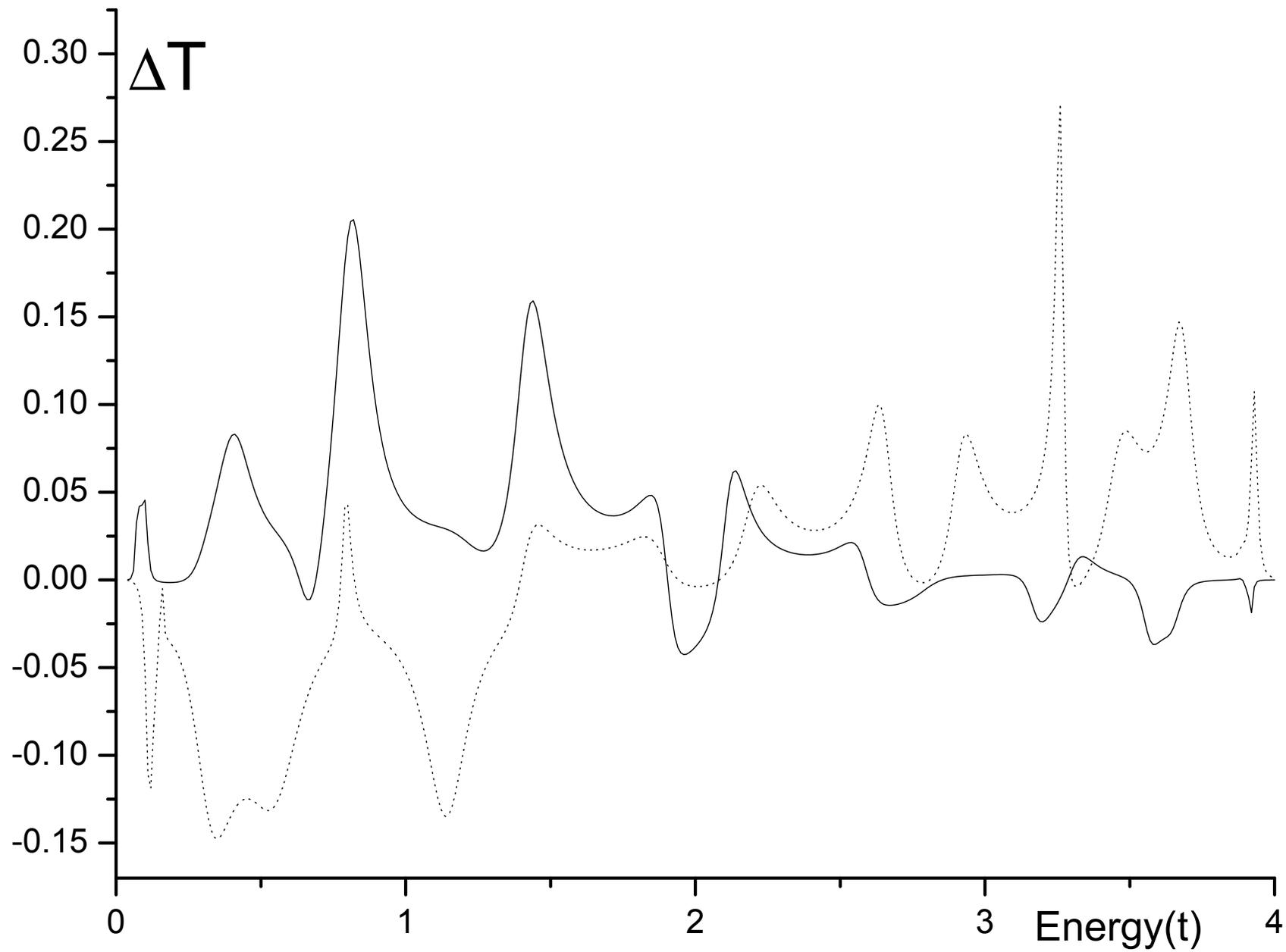
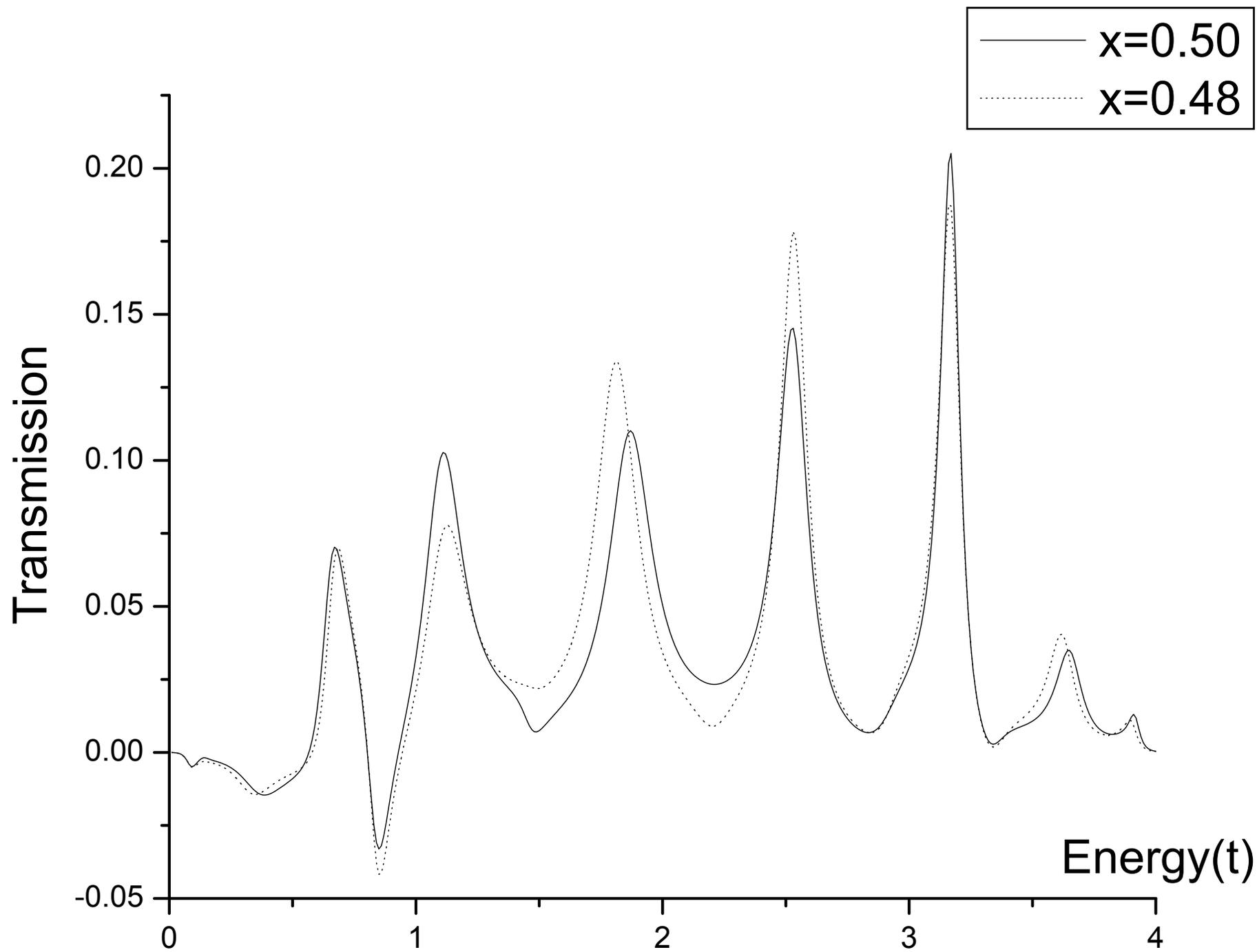
\end{document}